\documentclass[11pt]{article}
\usepackage{amsmath,amssymb,epsfig,bm}

\date{\empty}

\begin{document}

\title{\bf Relaxing the limits on inflationary magnetogenesis}

\author{Christos G. Tsagas\\ {\small Section of Astrophysics, Astronomy and Mechanics, Department of Physics}\\ {\small Aristotle University of Thessaloniki, Thessaloniki 54124, Greece}}

\maketitle

\begin{abstract}
Inflation has long been thought as the best way of producing primordial large-scale magnetic fields. To achieve fields strong enough to seed the galactic dynamo, most of the mechanisms operate outside conventional electromagnetic theory. The latter is typically restored after the end of the de Sitter phase. Breaking away from standard electromagnetism can lead to substantially stronger magnetic fields by the end of inflation, thus compensating for the their subsequent adiabatic depletion. We argue that the drastic magnetic enhancement during the de Sitter era may no longer be necessary because, contrary to the widespread perception, superhorizon-sized magnetic fields decay at a slower pace after inflation. The principle behind this claim is causality, which confines the post-inflationary electric currents inside the horizon. Without the currents there can be no magnetic-flux freezing on super-Hubble lengths. There, the magnetic decay-rate slows down, thus making it much easier to produce primordial fields of astrophysical interest. To quantify this qualitative statement, one can start from the current galactic-dynamo requirements and `reverse engineer' the magnetic strengths needed at the end of inflation, in order to produce astrophysically relevant residual seeds today. Our results suggest that, depending on the magnetic scale, mechanisms of inflationary magnetogenesis generating fields stronger than $10^{17}$~G by the end of the de Sitter phase, could successfully seed the galactic dynamo at present.
\end{abstract}\vspace{10mm}

Magnetic ($B$) fields appear everywhere in the universe~\cite{K}, but their origin remains an open issue~\cite{GR}. Inflation has a `natural' way of producing large-scale magnetic fields. However, to seed the galactic dynamo, the current strength of the field should lie between $10^{-22}$ and $10^{-12}$~Gauss, depending on the efficiency of the dynamo amplification~\cite{Kr}. Most mechanisms achieve such magnitudes by going beyond conventional electromagnetic theory during inflation~\cite{GR}. This can considerably increase the strength of the $B$-field by the end of the de Sitter phase and thus produce seeds that meet the galactic-dynamo requirements today. Maxwellian electromagnetism is restored after inflation and the magnetic flux is assumed to remain conserved for the rest of the field's evolution. In other words, the $B$-field is allowed to decay adiabatically (i.e.~$B\propto a^{-2}$, where $a$ is the cosmological scale factor) from the end of inflation to the present.

Assuming magnetic-flux conservation (i.e.~that $B\propto a^{-2}$) on all scales, means applying the ideal magnetohydrodynamic (MHD) approximation both inside and outside the particle horizon. After inflation, the latter essentially equals the Hubble radius. However, the MHD limit is the macroscopic outcome of causal microphysical processes that operate within the horizon. Extending the ideal-MHD approximation on super-Hubble scales, a priori accepts the presence of highly conductive electric currents with super-Hubble correlations. Recall that it is the currents which eliminate the electric fields and then freeze the accompanying magnetic fields into the cosmic fluid. These electric currents, however, are produced after inflation and their coherence length is always smaller than the horizon. Moreover, the processes of electric-field elimination and magnetic-flux freezing are not instantaneous but causal. Inside the horizon, the time required for the aforementioned processes to complete is typically much shorter than the expansion timescale. However, the time required to, say, freeze-in the superhorizon-sized magnetic fields is longer that the age of the universe. Put another way, the whole of the $B$-field must come into causal contact before it freezes-in. Hence, we cannot apply the ideal-MHD limit outside the Hubble radius without violating causality. All these mean that, even after inflation is over, on super-Hubble scales we are still dealing with the free magnetic fields left there from inflation.

The aforementioned causality claim was put forward in~\cite{BT}. An extensive discussion of this argument and of its potentially pivotal implications for cosmic magnetogenesis (conventional or not), namely that it allows for residual $B$-fields much stronger than those anticipated in the standard literature, was given in~\cite{T}. Very recently, the same causality claim and its implications were used to revisit the Ratra model of primordial magnetic generation~\cite{C}. Here, we will first outline how by appealing to causality one can slow down the post-inflationary adiabatic decay of superhorizon-sized magnetic fields on spatially flat Friedmann-Robertson-Walker (FRW) universes. We will then proceed to our main objective, which is to `reverse engineer' the magnetic strengths needed by the end of inflation, in order to produce seeds that lie within the galactic-dynamo requirements today. Our numerical estimates (summarised in TABLE~\ref{tab1}) are scale-dependent and show that, when causality is accounted for, the standard limits on inflationary magnetogenesis can relax considerably.

We begin by recalling that the galactic dynamo typically requires seeds with strengths in the range
\begin{equation}
10^{-22}~{\rm G}\lesssim B_0\lesssim 10^{-12}~{\rm G}\,,  \label{GDB0}
\end{equation}
where $B_0$ is the magnetic field today~\cite{Kr}. Moreover, these fields must also have a minimum (comoving) coherence length of approximately 10~Kpc. Magnetic fields weaker than the above given lower limit are very unlikely to sustain the dynamo.\footnote{Conventional inflationary magnetogenesis, where the $B$-field is allowed to decay adiabatically after the end of the de Sitter phase, leads to $B_0\sim10^{-53}$~G on scales close to $10$~Kpc~\cite{GR}. Note that this translates into $B_{DS}\sim10^{12}$~G at the end of inflation proper (see Eq.~(\ref{aBDS})). The aforementioned extreme weakness of the residual field has so far been the main reason for seeking solutions outside the realm of standard electromagnetic theory.} Those exceeding the upper limit, on the other hand, would probably lead to galactic fields stronger than the observed and perhaps violate cosmological constraints, like those set by primordial nucleosynthesis or by the cosmic microwave background isotropy (around $10^{-7}$~G and $10^{-9}$~G respectively -- e.g.~see~\cite{GR} and references therein).

Assuming that magnetic fields decay adiabatically throughout their post-inflationary evolution and keeping in mind that $a\propto T^{-1}$ always (where $T$ is the temperature of the universe), we find that $B_{RH}= B_0({T_{RH}/T_0})^2$ at the end of reheating. Then, at the end of the de Sitter phase, we have
\begin{equation}
B_{DS}\sim B_0\,{M^{8/3}\over T_0^2T_{RH}^{2/3}}\,,  \label{aBDS}
\end{equation}
where $M$ is the energy scale of the adopted inflation model. In deriving the above, we have used the fact that $\rho\propto a^{-3}$ throughout reheating (where $\rho$ is the density of the dominant matter component) and have set $\rho_{DS}\sim M^4$ and $\rho_{RH}\sim T_{RH}^4$ (with $M$, $T_0$ and $T_{RH}$ measured in GeV). Consequently, if magnetic fields were to decay adiabatically after inflation and still fulfill the galactic dynamo requirement today (i.e.~satisfy constraint (\ref{GDB0})), their value (in Gauss) at the end of the de Sitter phase must lie within the range
\begin{equation}
10^{-22}{M^{8/3}\over T_0^2T_{RH}^{2/3}}\lesssim B_{DS}\lesssim 10^{-12}{M^{8/3}\over T_0^2T_{RH}^{2/3}}\,.  \label{aBDScon1}
\end{equation}
Setting $T_0\sim10^{-13}$~GeV, $T_{RH}\sim10^{10}$~GeV and assuming GUT-scale inflation with $M\sim10^{17}$~GeV, the above translates into
\begin{equation}
10^{43}~{\rm G} \lesssim B_{DS}\lesssim 10^{53}~{\rm G}\,.  \label{aBDScon2}
\end{equation}

This stringent constraint, which is generally very hard to satisfy, applies  to magnetic fields that have been frozen into the cosmic medium from the end of inflation to the present. However, causality guarantees that there are no coherent electric currents with superhorizon correlations, which in turn implies that there is no magnetic flux-freezing on super-Hubble scales. Put another way, the adiabatic decay-law in not guaranteed on superhorizon lengths~\cite{BT,T}. Let us consider the implications of this claim, referring the reader to~\cite{T} for further discussion. In the absence of electric currents and on a spatially flat FRW background, the magnetic ($B_a$) component of the Maxwell field obeys the linear wave-like formula~\cite{GR}
\begin{equation}
\mathcal{B}^{\prime\prime}_a- a^2{\rm D}^2\mathcal{B}_a= 0\,.  \label{cBa''}
\end{equation}
Here, $\mathcal{B}_a=a^2B_a$ is the rescaled magnetic field, primes denote differentiation with respect to the conformal time ($\eta$) and ${\rm D}^2={\rm D}^a{\rm D}_a$ is the 3-dimensional Laplacian.\footnote{Inside the horizon the currents have the time to freeze the magnetic fields into the matter and the ideal-MHD holds. There, $B$-fields no longer obey Eq.~(\ref{cBa''}) but decay adiabatically (i.e.~$B\propto a^{-2}$), irrespective of the type of matter that fills the universe.} After harmonically decomposing $\mathcal{B}_a$, Eq.~(\ref{cBa''}) accepts the solution
\begin{equation}
\mathcal{B}_{(n)}= \mathcal{C}_1\cos(n\eta)+ \mathcal{C}_2\sin(n\eta)\,,  \label{cBn}
\end{equation}
where $n$ (with $n\geq0$) is the comoving wavenumber of the magnetic mode and $\mathcal{C}_{1,2}$ are the integration constants. Here, we will only be interested in superhorizon scales, where $n\eta\ll1$. There, a simple Taylor expansion reduces (\ref{cBn}) to the power law
\begin{equation}
\mathcal{B}= a^2B= \mathcal{C}_1+ \mathcal{C}_2n\eta\,.  \label{Bn}
\end{equation}
The second mode on the right-hand side of the above ensures that post-inflationary $B$-fields do not necessarily decay adiabatically, as long as they are outside the Hubble radius. Note that the fact that $n\eta\ll1$ does not guarantee that the second mode of solution (\ref{Bn}) is negligible. For example, when the initial conditions lead to $\mathcal{C}_2\gg\mathcal{C}_1$, the second mode could play the key role. One therefore needs to evaluate the integration constants first, which is what we are going to do next.

Throughout reheating and during the dust era $a\propto\eta^2$. Then, $\mathcal{C}_1=-(3B_*+\eta_*B_*^{\prime})a_*^2$ and $\mathcal{C}_2=(4B_*+\eta_*B_*^{\prime})a_*^2/n\eta_*$, with the ``$*$''-suffix indicating the transition moment from one cosmological epoch to the next. Since $n\eta_*\ll1$, we deduce that $C_2\gg\mathcal{C}_1$ (unless $4B_*+\eta_*B_*^{\prime}=0$), which shows why we cannot a priori ignore the second mode of (\ref{Bn}). Then, we arrive at
\begin{equation}
B= -\left(3B_*^{+}+\eta_*^{+}B_*^{\prime\;+}\right) \left({a_*^{+}\over a}\right)^2+ \left(4B_*^{+}+\eta_*^{+}B_*^{\prime\;+}\right) \left({a_*^{+}\over a}\right)^{3/2}\,,  \label{RH-dB1}
\end{equation}
with the ``+''-superscript marking the start of the associated era. Similarly, evaluating $\mathcal{C}_1$ and $\mathcal{C}_2$ in the radiation era, when $a\propto\eta$, recasts Eq.~(\ref{Bn}) into
\begin{equation}
B= -\left(B_*^{+}+\eta_*^{+}B_*^{\prime\;+}\right) \left({a_*^{+}\over a}\right)^2+ \left(2B_*^{+}+\eta_*^{+}B_*^{\prime\;+}\right) \left({a_*^{+}\over a}\right)\,.  \label{rB}
\end{equation}
Solutions (\ref{RH-dB1}) and (\ref{rB}) contain modes that decay slower than the adiabatic rate. More specifically, $B\propto a^{-3/2}$ during reheating and after equipartition, while $B\propto a^{-1}$ in the radiation epoch. Thus, as long as it remains outside the Hubble radius, the $B$-field is superadiabatically amplified throughout its post-inflationary evolution. Once back inside the horizon, on the other hand, the adiabatic decay-law is `reinstated'.

The integration constants of (\ref{RH-dB1}) and (\ref{rB}) depend on the initial conditions. At the start of reheating, these are decided by the magnetic evolution in the de Sitter phase and by the nature of the transition from inflation to reheating. The same principles also apply to the radiation and the dust epochs. As long as the initial conditions allow the slowly decaying modes of (\ref{RH-dB1}) and (\ref{rB}) to survive, superhorizon-sized magnetic fields will be superadiabatically amplified. This happens in non-conventional scenarios of cosmic magnetogenesis.\footnote{Initial conditions allowing the slowly decaying modes of solutions (\ref{RH-dB1}) and (\ref{rB}) to survive are possible in conventional scenarios of cosmic magnetogenesis as  well~\cite{T}.} There, $B$-fields typically decay as $B\propto a^{-m}$ (with $0\leq m<2$, $a\propto-1/\eta$ and $\eta<0$) all along the de Sitter expansion (e.g.~see~\cite{S}). This implies $B_*^{\prime\;-}= mB_*^{-}/\eta_*^{-}$ at the end of inflation. Setting $\eta_*^{+}=-\eta_*^{-}$, $B_*^{+}=B_*^{-}$ and $B_*^{\prime\;+}=B_*^{\prime\;-}$ on either side of the transition hypersurface (recall that $\eta_*^-<0$ and $\eta_*^+>0$), gives $B_*^{\prime\;+}= -mB_*^{+}/\eta_*^{+}$ at the start of reheating. With these initial conditions, solution (\ref{RH-dB1}) reads
\begin{eqnarray}
B= -(3-m)B_*^{+}\left({a_*^+\over a}\right)^2+ (4-m)B_*^{+}\left({a_*^+\over a}\right)^{3/2}\hspace{-6pt}\,.  \label{RH-dB2}
\end{eqnarray}
Since $0\leq m<2$, the above confirms that $B\propto a^{-3/2}$ and a superadiabatic amplification for the field throughout the reheating era. Similarly, one can demonstrate that the magnetic (superadiabatic) amplification carries on during the radiation and the dust epochs, as long as the $B$-field is still outside the horizon~\cite{T}.

All these mean that magnetic fields generated by non-conventional mechanisms during inflation can have much stronger residual values than those anticipated. Then, it might not be necessary to produce very strong $B$-fields by the end of inflation to achieve astrophysically interesting values today. In other words, the standard constraints (\ref{aBDScon1}) and (\ref{aBDScon2}) can relax considerably. To estimate how much is our main objective and in order to do so we need to know the scale of the magnetic field. This determines the time of horizon entry, which marks the transition from superadiabatic amplification (outside the Hubble radius) to adiabatic decay (inside the Hubble length). The longer a magnetic mode stays outside the horizon the stronger its amplifucation, in  which case (\ref{aBDScon1}) and (\ref{aBDScon2}) can relax considerably. Next, we will re-evaluate these constraints for fields that enter the horizon (i) in the radiation era and (ii) during the dust epoch.\\

(i) Suppose that a magnetic mode crosses inside the Hubble radius some time in the radiation epoch. Then onwards $B\propto a^{-2}$, which means that $B_{HC}= B_0(T_{HC}/T_0)^2$ at horizon crossing.
Earlier in the radiation era and also during reheating, the mode is outside the Hubble scale and decays as $B\propto a^{-1}$ and $B\propto a^{-3/2}$ respectively. With these in mind, a straightforward calculation leads to
\begin{equation}
B_{DS}\sim B_0\,{T_{HC}M^2\over T_0^2T_{RH}}\,,  \label{BDS1}
\end{equation}
at the end of the de Sitter expansion (recall that $\rho_{DS}\sim M^4$ and $\rho_{RH}\sim T_{RH}^4$ in natural units). Therefore, to achieve astrophysically relevant residual values today, the magnetic strength (in Gauss) must lie within the range
\begin{equation}
10^{-22}{T_{HC}M^2\over T_0^2T_{RH}}\lesssim B_{DS}\lesssim 10^{-12}{T_{HC}M^2\over T_0^2T_{RH}}\,,  \label{BDScon1}
\end{equation}
at the end of inflation proper.

Let us evaluate the above in a particular case. Consider a $B$-field with current scale $\lambda_0\sim10$~Kpc, which is the minimum required for the galactic dynamo to work. Since $\lambda\propto a$ always and $\lambda_H\propto a^{3/2}$ during the dust era, we have $(\lambda_H/\lambda)_{EQ}=(\lambda_H/\lambda)_0 (T_0/T_{EQ})^{1/2}\sim10^3$, when $T_0\sim10^{-13}$~GeV, $T_{EQ}\sim10^{-9}$~GeV and $(\lambda_H)_0\sim10^3$~Mpc. Therefore, a mode of approximately 10~Kpc at present entered the horizon before equipartition. In the radiation era $\lambda_H\propto a^2$, which means that $(\lambda_H/\lambda)_{HC}=(\lambda_H/\lambda)_{EQ} (T_{EQ}/T_{HC})$. Then, the aforementioned magnetic mode crossed inside the Hubble radius at $T_{HC}\sim10^{-6}$~GeV. Substituting the above into (\ref{BDScon1}), setting $T_{RH}\sim10^{10}$~GeV and assuming GUT-scale inflation with $M\sim10^{17}$~GeV gives
\begin{equation}
10^{22}~{\rm G} \lesssim B_{DS}\lesssim 10^{32}~{\rm G}\,.  \label{BDScon2}
\end{equation}
Therefore, inflationary $B$-fields with current size close to 10~Kpc can seed the galactic dynamo without satisfying the `adiabatic' constraint (\ref{aBDScon2}), but the drastically more relaxed limits given above (see also TABLE~\ref{tab1}).\\

(ii) Consider magnetic fields entering the horizon after matter-radiation equality. If $B_0$ is the magnetic strength today, we have $B_{HC}= B_0(T_{HC}/T_0)^2$ at horizon crossing. Then, given that $B\propto a^{-3/2}$ on super-Hubble scales as long as dust dominates, we find that
\begin{equation}
B_{EQ}= B_0{T_{HC}^{1/2}T_{EQ}^{3/2}\over T_0^2}\,,  \label{BEQ1}
\end{equation}
at equilibrium. Proceeding exactly as before (see case (i) previously), we arrive at
\begin{equation}
B_{DS}\sim B_0{T_{HC}^{1/2}T_{EQ}^{1/2}M^2\over T_0^2T_{RH}}\,,  \label{BDS2}
\end{equation}
at the end of inflation. This magnetic field will satisfy the dynamo requirements today if
\begin{equation}
10^{-22}{T_{HC}^{1/2}T_{EQ}^{1/2}M^2\over T_0^2T_{RH}}\lesssim B_{DS}\lesssim 10^{-12}{T_{HC}^{1/2}T_{EQ}^{1/2}M^2\over T_0^2T_{RH}}\,.  \label{BDScon3}
\end{equation}
Wavelengths entering the horizon at $T_{HC}>10^{-6}$~GeV have current sizes greater that 10~Kpc (see case (i) above) and fulfill the dynamo's scale-requirement. Also, $B$-fields on such scales have stayed outside the Hubble radius longer and therefore their residual values are stronger. This means that they can have smaller magnitudes at the end of inflation and still seed the galactic dynamo. For example, according to (\ref{BDScon3}), a magnetic field entering the horizon today (i.e.~with $T_{HC}\sim T_0\sim10^{-13}$~GeV) will work if it satisfied the constraint
\begin{equation}
10^{17}~{\rm G} \lesssim B_{DS}\lesssim 10^{27}~{\rm G}\,,  \label{BDScon4}
\end{equation}
at the end of the de Sitter expansion (see also TABLE~\ref{tab1}). The above limits are clearly more relaxed than those of (\ref{BDScon2}) and far more relaxed than those of the adiabatic scenario (see constraint (\ref{aBDScon2}) earlier).

\begin{table}
\caption{The strength-range of inflationary magnetic fields, measured at the end of the de Sitter phase, capable of seeding the galactic dynamo today (compare to the standard -- `adiabatic' -- range given in Eq.~(\ref{aBDScon2})). The first row corresponds to $B$-fields with the minimum required scale for the dynamo to work. The third and forth rows refer to magnetic modes that crossed inside the horizon at recombination and at present respectively. In all cases $M\sim10^{17}$~GeV and $T_{RH}\sim10^{10}$~GeV.}
\begin{center}\begin{tabular}{cccccc}
\hline \hline & \hspace{2mm} $\lambda_0~({\rm Mpc})$ \hspace{3mm} & $T_{HC}~({\rm GeV})$ \hspace{3mm} & $B_{DS}~({\rm G})$ &\\ \hline \hline & $\begin{array}{c}  10^{-2} \\ 1 \\ 10^{3/2} \\ 10^3 \end{array}$ & $\begin{array}{c} \hspace{-4mm} 10^{-6} \\ \hspace{-4mm} 10^{-8} \\ \hspace{-4mm} 10^{-10} \\ \hspace{-4mm} 10^{-13} \end{array}$ & $\begin{array}{c} 10^{22}\lesssim B_{DS}\lesssim 10^{32} \\ 10^{20}\lesssim B_{DS}\lesssim 10^{30} \\ 10^{18}\lesssim B_{DS}\lesssim 10^{28} \\ 10^{17}\lesssim B_{DS}\lesssim10^{27} \end{array}$\\ [2.5truemm] \hline \hline
\end{tabular}\end{center}\label{tab1}\vspace{-0.5truecm}
\end{table}

To achieve current magnetic strengths that satisfy the `adiabatic' constraints (\ref{aBDScon1}) and (\ref{aBDScon2}), non-conventional mechanisms of inflationary magnetogenesis enhance their $B$-fields substantially in the de Sitter phase. Producing strong
magnetic fields during inflation, however, is not a problem-free exercise. One issue is the so-called backreaction effect, where the $B$-field becomes strong enough to interfere with the dynamics of the expansion~\cite{DMR,KSW}.

As long as they remain outside the Hubble radius, however, the aforementioned fields decay at a pace slower than the adiabatic for the rest of their evolution. Then, strong amplification during inflation is not only unnecessary, but it can be problematic as well. Instead, a relatively weak enhancement in the de Sitter phase could suffice. To illustrate these points consider a scenario leading to $B$-fields as strong as $\sim10^{46}$~G at the end of inflation on all scales~\cite{DMR,KSW}. Magnetic fields of such magnitude satisfy the dynamo requirements today, even if we assume that they have been decaying adiabatically throughout their post-inflationary life (see constraint (\ref{aBDScon2})). However, causality ensures that these fields have been decaying at a considerably slower pace for as long as they stayed outside the Hubble radius. A magnetic mode that enters the horizon today, for example, will have current strength close to $10^7$~G (see Eq.~(\ref{BDS2})), which is clearly at odds with the observations. On the other hand, suppose that the generated field has strength close to $10^{22}$~G at the end of inflation and current size around 1~Mpc~\cite{KSW}. Such fields cannot seed the dynamo today, if the adiabatic decay-law is imposed on all scales after inflation. Following TABLE~\ref{tab1}, however, magnetic fields of Mpc-size at present and magnitude close to $10^{22}$~G at the end of the de Sitter phase can sustain the dynamo. This happens because the aforementioned fields have remained outside the Hubble radius until late into the radiation era and crossed inside at $T_{HC}\sim10^{-8}$~GeV  (see TABLE~\ref{tab1}). As long as they stayed outside the horizon, these $B$-fields were superadiabatically amplified. Consequently, their residual strength increases to $10^{-20}$~G (see Eq.~(\ref{BDS1})), which can seed the dynamo.

The origin of cosmic magnetism remains unresolved and, over the years, mechanisms of inflationary magnetogenesis that operate outside standard electromagnetic theory have become increasingly popular. The aim is to produce substantially strong $B$-fields already by the end of the de Sitter phase. Then, although classical electromagnetism is restored after inflation, the residual $B$-field is still capable of seeding the galactic dynamo.

A common and key assumption in all these scenarios is that post-inflationary $B$-fields decay adiabatically, even on superhorizon scales. This, however, violates causality, which confines the processes of electric-current generation and magnetic-flux freezing within the causal horizon. Put another way, the ideal-MHD limit and the adiabatic $B\propto a^{-2}$-law do not apply on scales larger than the Hubble radius. There, $B$-fields decay slower throughout their post-inflationary life. Consequently, the final magnetic strengths produced by the majority of these non-conventional mechanisms are much larger than anticipated. In fact, most of the $B$-fields produced in these scenarios are so strong by the end of the de Sitter phase that their current strengths are well in excess of those measured in the galaxies, or of those allowed by the observations.

Every cloud has a  silver lining however. The slow decay of superhorizon-sized $B$-fields after inflation, means that a relatively mild amplification during the de Sitter phase can suffice. Put another way, inflationary magnetic seeds may not need to satisfy the stringent requirements set by constraint (\ref{aBDScon2}), in order to be of astrophysical relevance. Indeed, starting from the current galactic-dynamo requirements and by `reverse engineering' the magnetic strengths, we found that mechanisms of magnetic generation producing fields stronger than $10^{17}$~G at the end of inflation can seed the dynamo today (see TABLE~\ref{tab1}). These new limits mainly target the non-conventional scenarios of inflationary magnetogenesis, since their conventional counterparts typically produce considerably weaker $B$-fields. In~\cite{KSW}, for example, the authors discuss a (non-conventional) scenario that produces magnetic fields close to $10^{22}$~G at the end of the de Sitter expansion, with current size around 1~Mpc. Assuming adiabatic decay on all scales after inflation, brings the magnitude of the aforementioned field to $10^{-43}$~G at present, which cannot seed the dynamo. In view of our revised limits, however, the aforementioned field can seed the dynamo (see TABLE~\ref{tab1}, second row). This happens because causality and the absence of superhorizon-sized electric currents has slowed down the post-inflationary evolution of this field and thus increased the previously quoted strength to $10^{-20}$~G. The latter lies within the typical galactic-dynamo requirements.

\end{document}